\documentstyle[prl,aps,epsf,floats]{revtex}
\begin{document}
\twocolumn[
\hsize\textwidth\columnwidth\hsize\csname@twocolumnfalse\endcsname
\draft
]
\noindent{\bf 
Comment on "History Effects and Phase Diagram near the Lower Critical Point 
in ${\rm Y} {\rm Ba}_2 {\rm Cu}_3 {\rm O}_{7-\delta}$ Single Crystals":}

\bigskip

\ A lower critical point $B_{\rm lmc}$ of the first order transition 
line $B_m(T)$ signaled by a magnetization jump often appears in 
${\rm Y Ba}_2 {\rm Cu}_3 {\rm O}_{7-\delta}$ (YBCO) 
at a magnetic field in tesla range, for example, as a result of 
enhanced {\it extrinsic} point disorder\cite{Kwok}. 
Recently, Zhukov et al.\cite{Zhu} have found in their magnetization data 
the presence of $B_{\rm lmc}$ and of a history effect 
in {\it lower} fields than $B_{\rm lmc}$. To try to explain $B_{\rm lmc}$ 
{\it increasing}\cite{Zhu,Nishi} with decreasing oxygen-deficiency $\delta$ 
acting as an extrinsic pinning disorder, 
they have invoked an intrinsic pinning 
disorder, of electronic origin, increasing with decreasing $\delta$ by 
assuming $B_{\rm lmc}$ as well as the upper critical point $B_{\rm umc}$ to be 
induced by some (static) disorder. However, since {\it both} $B_{\rm umc}$ 
and $B_{\rm lmc}$ decrease with increasing $\delta$ \cite{Nishi}, 
a comprehensive 
understanding of the phase diagram based on the argument in ref.2 would 
require such an unacceptable assumption that $B_{\rm lmc}$ is controlled by 
the intrinsic disorder, 
while $B_{\rm umc}$ is mainly determined by the extrinsic one. 

It is well accepted at present that the macroscopic properties 
in vortex states of real type II superconductors should be described based on 
the Ginzburg-Landau (GL) model characterized by three material 
parameters, i.e., the fluctuation strength, anisotropy (dimensionality), and 
pinning strength. Since a single pinning strength describes possible 
point-like disorders altogether in GL theory, the doping dependences of $B_{\rm lmc}$ and 
$B_{\rm umc}$ cannot be understood consistently even if artificially 
introducing\cite{Zhu} a couple of different pinning disorders. Below, we 
briefly point out that the dependences of the critical points on $\delta$ 
\cite{Zhu,Nishi} and on an {\it extrinsic} pinning disorder\cite{Kwok} can be 
explained altogether within the GL theory\cite{RI1,RI2} if taking account of 
the fact that the mass anisotropy $\Gamma$ and the fluctuation strength 
($\propto \lambda^2(T)$) increase with increasing $\delta$, 
where $\lambda(T)$ is the in-plane penetration depth. 
First, we note that $B_{\rm umc}$ is generally believed\cite{RI1} 
to decrease with increasing an effective pinning strength 
$D_{p, {\rm eff}}$ (a relative strength of point disorder to the fluctuation 
strength) which is effectively enhanced by increasing fields. 
This statement is valid particularly in type II limit where the magnetic screening, important near $H_{c1}$, is neglected. Assuming the type II limit in 
the field range far above $H_{c1}$ where the two critical points were 
detected\cite{Zhu,Nishi} in YBCO is safely valid. 
Rather, in our theory\cite{RI2}, 
the appearance of $B_{\rm lmc}$ is based on the fact\cite{RI1} that, 
in layered systems, the vortex-glass (VG) transition curve $B_{\rm VG}(T)$ in 
type II limit can lie above the $B_m(T)$-line (defined in clean limit) in 
lower fields only because of a difference in their $T$-dependences. 
Since the Ohmic resistance vanishes at $B_{\rm VG}(T)$ possibly corresponding to $B_p(T)$ in ref.2, 
the $B_m(T)$-line in $B < B_{\rm lmc}$ 
lies within a glass phase. Namely, the superconducting transition 
in $B < B_{\rm lmc}$ is a continuous VG transition, and $B_m(T)$ is 
not realized there as the {\it thermal} first order transition 
signaled by a jump of reversible magnetization. Up to the lowest order in the 
bare pinning strength, the resulting lower critical point is estimated as 
$$B_{\rm lmc} \simeq c_L H_{c2}(0) \, \biggl({{\phi_0^2 \, d} \over {16 \pi^2 \lambda^2(0) k_{\rm B}T_c}} \biggr) \, {{{\xi_0^4} \over {d^4 \Gamma^2}}} \, D_{p, {\rm eff}}^6, 
\eqno(1)$$ 
where $d$ the layer spacing, $\xi_0$ the in-plane coherence length, 
and $c_L$ is a constant taking a value between $10^{-2}$ 
and $10^{-3}$. The inverse of parameter combination in the bracket measures the fluctuation strength per layer. In contrast to $B_{\rm umc}$, 
the primary origin of $B_{\rm lmc}$ decreasing\cite{Zhu,Nishi} with increasing $\delta$ is, according to eq.(1), its remarkable dependence on $\Gamma$ and $\lambda(0)$ both of which increase with increasing $\delta$. Although it is not straightforward to predict a $B_{\rm lmc}$'s sample dependence because it 
depends competitively on all of the three material parameters, nevertheless 
we note that, by assuming $\Gamma \leq 8$ for the fully-oxidized case 
and, for brevity, other parameters to have no doping dependences, 
$B_{\rm lmc}$ may stay in tesla range on the overdoped side. Eq.(1) may be 
insufficient in {\it quantitatively} explaining lower $B_{\rm lmc}$ values in 
optimally doped samples where the higher Landau level fluctuations neglected in obtaining eq.(1) are no longer negligible 
below 1.7 (T) \cite{RI3}. 
Nevertheless, the $\Gamma$-dependence of eq.(1) is qualitatively 
consistent not only with the absence (in accessible fields) of $B_{\rm lmc}$ in BSCCO and underdoped YBCO with quite large $\Gamma$-values but also with 
a $\delta$-dependence of $B_{\rm lmc}$ suggested from optimally-doped and 
three overdoped YBCO samples\cite{Nishi}. 

Further, according to recent works\cite{Sweden}, a $B_{\rm lmc}$ of several 
tesla is present in the parallel field case even in underdoped YBCO. This fact 
is compatible not with the interpretation invoking an intrinsic 
disorder\cite{Zhu} negligible in underdoped YBCO but only with the present 
theory based on the GL model with extrinsic disorder\cite{para}. 

We are grateful to T. Nishizaki for informative discussion. 

\vspace{0.3truecm}
\bigskip

\noindent Ryusuke Ikeda \\
Department of Physics, \\
Kyoto University, Kyoto 606-8502, Japan

\end{document}